\documentclass[prl,twocolumn,showpacs,amsmath,amssymb,superscriptaddress]{revtex4-1}
\usepackage[latin9]{inputenc}
\usepackage{float}
\usepackage{amsmath}
\usepackage{amssymb}
\usepackage{graphicx}
\usepackage{color}
\usepackage{esint}
\usepackage[author={Nils},color={0 .13 .26},open=true,width=2]{pdfcomment}
\usepackage{rotating}\usepackage{braket}\usepackage{amsthm}\usepackage{bbm}\newcommand{\ens}[0]{\ensuremath}
\newcommand{\iE}[0]{\ens{\mathrm{i}}}

\usepackage{subfigure}
\makeatother

\begin{document}

  \title{Time-Reversal-Symmetric Single-Photon Wave Packets for Free-Space
Quantum Communication}

  \author{N. Trautmann}
  
   \affiliation{Institut f\"ur Angewandte Physik, Technische Universit\"at Darmstadt, 64289 Darmstadt,
  Germany}
  
   \author{G. Alber}

 \affiliation{Institut f\"ur Angewandte Physik, Technische Universit\"at Darmstadt, 64289 Darmstadt,
  Germany}
  
  \author{G. S. Agarwal}
  \affiliation{Department of Physics, Oklahoma State University, Stillwater, Oklahoma 74078, USA}
  
  \author{G. Leuchs}
  \affiliation{Max-Planck-Institut f\"ur die Physik des Lichts,
    G\"{u}nther-Scharowsky-Stra{\ss}e 1, Bau 24, 91058 Erlangen,
    Germany} 
  \affiliation{Department f\"{u}r Physik, Universit\"{a}t Erlangen-N\"{u}rnberg,
  Staudtstra{\ss}e 7, Bau 2, 91058 Erlangen, Germany}
  
  \date{\today}
  \begin{abstract}
  Readout and retrieval processes are proposed for efficient, high-fidelity quantum state transfer between
  a matter qubit, encoded in the level structure of a single atom or ion, and a 
  photonic qubit, encoded in a time-reversal-symmetric single-photon wave packet. 
  They are based on controlling spontaneous photon emission and absorption of a matter qubit on demand in free
  space by stimulated Raman adiabatic passage. 
  As these processes do not involve mode selection by high-finesse cavities or photon transport through optical fibers, they offer interesting perspectives as basic building blocks for free-space quantum-communication protocols.
  \end{abstract}
  

\maketitle
High-fidelity quantum state transfer between single photons acting as ``flying qubits'' and matter qubits, such as atoms or ions, acting as ``stationary'' memory ``qubits''
is a crucial process for quantum technological applications. It is an important elementary readout and retrieval process in quantum-communication protocols \cite{Nielsen} in which quantum information exchange between distant stationary matter qubits is achieved by photon transport through optical fibers or through free-space channels. Whereas fiber based quantum communication offers advantages for local networks \cite{Kimble}
free-space implementations are of special interest for the realization of a future worldwide satellite-based quantum-communication network
\cite{ursin2007entanglement}.

Motivated by its significance for quantum communication the realization of efficient light matter coupling in free space has been subject of recent experimental \cite{fischer2013,lindlein2007new,tey2008strong}
and theoretical \cite{Stobinska2009}
investigations.
Although efficient, high-fidelity quantum state transfer from a matter qubit to a photonic qubit can be achieved by spontaneous photon emission, realizing
the reverse process 
in free space is a formidable experimental challenge.
Exploiting time-reversal symmetry it
has been shown that almost perfect absorption of a single photon in free space 
is possible provided the single-photon wave packet
has an exponentially growing temporal envelope  \cite{Stobinska2009}.
An interesting probabilistic method for generating and shaping a single-photon wave packet with an exponentially rising envelope has been developed recently
\cite{kolchin2008electro,liu2014efficiently}. 
The inherent probabilistic nature of the procedure stems from the usage of a photon source based on spontaneous four-wave mixing \cite{balic2005generation} and the
shaping of the wave packet by using electro-optical amplitude modulation.
However, the probabilistic nature of these processes is a disadvantage for
applications concerning readout and retrieval processes of quantum information and procedures capable of performing such tasks on demand in a deterministic way are favorable.

In fiber and cavity based quantum-communication schemes a proposal to overcome the obstacles
of probabilistic photon generation and probabilistic wave packet shaping has been
developed by Cirac. et al. \cite{Cirac1997} and has been implemented experimentally by Ritter et al. \cite{Ritter}, recently.
In this experiment a laser pulse
controls the interaction of a single trapped atom with the radiation field inside a high-finesse cavity.
Exploiting the extreme cavity-induced mode selection of the high-finesse cavity and the resulting vacuum Rabi oscillations
governing the coherent spontaneous photon emission and absorption processes, a matter qubit can be converted efficiently on demand to a single-photonic qubit prepared in a time-reversal-symmetric wave packet state.
Due to this symmetry of
the wave packet
the quantum information stored in this single photon can be 
retrieved with high fidelity after transmission through an optical fiber and stored in another matter qubit.

However, this efficient readout and retrieval procedure
is not suitable for free-space quantum-communication protocols which do not involve any strong mode selection mechanism
because in free space the spontaneous photon emission process is of a considerably different nature. It is no longer governed  by coherent vacuum Rabi oscillations but by an
approximate exponential decay of the matter qubit. This decay reflects the extreme multimode aspects of the spontaneous emission process.
In view of these differences the natural question arises whether it is possible to design efficient, high-fidelity readout and retrieval procedures for photon-mediated quantum information transfer in free space on demand.

\begin{figure}[H]
\begin{center} \includegraphics[scale=0.5]{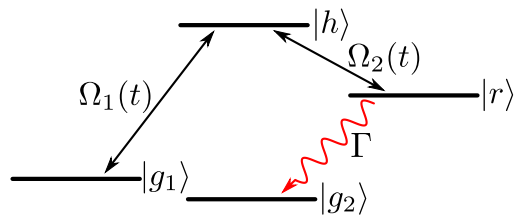}\end{center} 
\caption{(color online). Basic atomic level scheme  as described in the text.\label{fig:level_scheme}
}
\end{figure}

In the following it is demonstrated that 
such procedures
can be realized by appropriately controlling the spontaneous emission and absorption of a single photon by a matter qubit
with the help of stimulated Raman adiabatic passage (STIRAP) \cite{bergmann1998coherent}.
Thereby, instead of suppressing the multimode aspects of the matter-field
interaction
by a high-finesse cavity these features are exploited for
controlling 
the generation and absorption of a time-reversal-symmetric single-photon wave packet by a material qubit.
Recent experiments \cite{Maiwald,fischer2013} demonstrate that
in free-space scenarios material qubits can be trapped in the foci of parabolic cavities thus enabling efficient and well directed photon emission and absorption.
Combining these trapping techniques with
these readout and retrieval processes 
efficient free-space quantum-communication protocols can be designed for high-fidelity photon-mediated quantum state transfer between distant matter qubits.
\begin{figure}[H]
\begin{center} \includegraphics[scale=0.65]{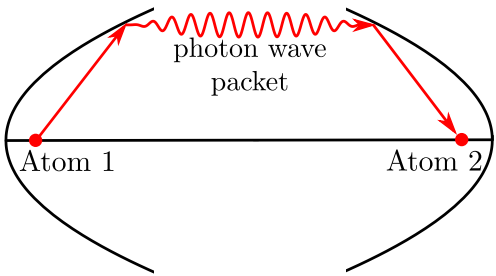}\end{center} 
\caption{(color online). Schematic representation of a free-space quantum state transfer protocol as described in the text.
\label{fig:Setup}}
\end{figure}
We consider a physical setup as depicted schematically in 
Fig.\,\ref{fig:Setup}. 
Two atoms or ions are trapped in the focal points of two distant opposite parabolic cavities  
whose focal length is  large and whose surface roughness is small compared with typical optical wavelengths.
The relevant atomic level structure is depicted in Fig.\,\ref{fig:level_scheme}.
The states $|g_{1}\rangle$ and $|g_{2}\rangle$ are radiatively stable or metastable ground states
constituting a logical qubit. The state $|h\rangle$
is an ancillary level which is used to transfer coherently
population between states
$|g_{1}\rangle$ and $|r\rangle$
by a STIRAP process. The (classical) laser pulses performing the STIRAP process are
characterized by the time dependent
Rabi frequencies $\Omega_{1}(t)$ and $\Omega_{2}(t)$. 
Preparing initially a trapped atom in state $|g_{1}\rangle$
an appropriate STIRAP-controlled spontaneous decay process from
level $|r\rangle$ to the ground state $|g_{2}\rangle$ with the
spontaneous decay rate $\Gamma$ is capable of generating a time-reversal-symmetric single-photon
wave packet.
For this purpose it is important that state $|r\rangle$ can decay only to
state $|g_{2}\rangle$ and not to state $|g_{1}\rangle$ or to any other state. 
Preparing initially a trapped atom in state $|g_{2}\rangle$
this time-symmetric single-photon wave packet can be absorbed again almost perfectly
by an appropriately tailored STIRAP-assisted photon absorption process.
These two processes are the basic building blocks of our proposed high-fidelity readout and retrieval processes.

In the dipole and rotating wave approximation the interaction of an atom in the focus of a parabolic cavity with the radiation field
is described by the Hamiltonian
$\hat{H}=\hat{H}_{\text{atom}}+\hat{H}_{\text{field}}+\hat{H}_{\text{i}}\label{eq:Hamiltonian}$
 with the Hamiltonians of the free radiation field 
$\hat{H}_{\text{field}}=\hbar\sum_{i}\omega_{i}a_{i}^{\dagger}a_{i}$
and of the free atom 
$\hat{H}_{\text{atom}}=\hbar\omega_{g_{1}}|g_{1}\rangle\langle g_{1}|+\hbar\omega_{g_{2}}|g_{2}\rangle\langle g_{2}|
+\hbar\omega_{h}|h\rangle\langle h|+\hbar\omega_{r}|r\rangle\langle r|$.
The modes of the quantized radiation field inside a parabolic cavity are characterized by the orthonormal mode functions
$\mathbf{g}_{i}(\mathbf{x})$
with frequencies $\omega_{i}$ and by the corresponding photonic destruction and creation operators $\hat{a}_i$ and $\hat{a}^{\dagger}_i$.
Describing the laser fields inducing the STIRAP process classically the interaction between the atom and the radiation field is characterized by the Hamiltonian
\begin{eqnarray}
&\hat{H}_{\text{i}}=-\left(\hat{\mathbf{E}}_{\perp}^{+}(\mathbf{x}_{a})\cdot\mathbf{d}|g_{2}\rangle\langle r|+\text{H.c.}\right)\nonumber\\
&-\frac{\hbar}{2}\left(e^{\iE(\omega_h-\omega_{g_1}+\Delta_1)\left(t-t_{0}\right)}\Omega_{1}(t)|g_{1}\rangle\langle h|+\text{H.c.}\right)\nonumber\\
&-\frac{\hbar}{2}\left( e^{\iE(\omega_h-\omega_{r}+\Delta_2)\left(t-t_{0}\right)}\Omega_{2}(t)|r\rangle\langle h|+\text{H.c.}\right)\,.
\end{eqnarray}
The time dependent Rabi frequencies
$\Omega_{1}(t)$ and $\Omega_{2}(t)$ characterize the interaction of these laser fields with 
the atom and
$\Delta_{1}$, $\Delta_{2}$ are their detunings from resonance.
The positive frequency part of the
electric field operator 
$\hat{\mathbf{E}}_{\perp}^{+}(\mathbf{x})$
describing the quantized modes of the radiation field
is given by
\begin{eqnarray}
\hat{\mathbf{E}}_{\perp}^{+}(\mathbf{x})&=&
\left(\hat{\mathbf{E}}_{\perp}^{-}(\mathbf{x})\right)^{\dagger}=
\iE\sum_{i}\sqrt{\frac{\hbar\omega_{i}}{2\epsilon_{0}}}\mathbf{g}_{i}(\mathbf{x})
\hat{a}_{i}^{\dagger}.
\end{eqnarray}
The dipole
matrix element of the atomic transition $|r\rangle\leftrightarrow|g_{2}\rangle$ is denoted 
$\mathbf{d}$ and 
$\mathbf{x}_a$ is the position of the atom. 
It is assumed that the spontaneous decay of the state $|h\rangle$ is negligible
as this state is not populated significantly during the STIRAP process.
This makes the spontaneous decay process from the state $\ket{r}$ to the state $\ket{g_2}$ highly controllable and robust \cite{bergmann1998coherent,STIRAP_decay}.

In order to describe our high-fidelity readout and retrieval processes we focus on atom-field states of the form 
$\ket{\psi_{\text{initial}}}=\ket{\psi_{\text{atom}}}\ket{\psi_{\text{field}}}$
initially prepared at time $t=t_0$.  
In the readout process which involves the generation of a time-symmetric single-photon wave packet initially the atom is prepared in
state $\ket{g_{1}}$ and the radiation field in the vacuum
state. Analogously, in the retrieval process, which involves almost perfect absorption of this wave packet,  
initially the atom is prepared in state 
$\ket{g_{2}}$
and the radiation field in this one-photon wave packet state.
For these initial states the time dependent Schr\"odinger equation can be solved by the ansatz
\begin{eqnarray}
&\ket{\psi(t)}=\ket{\psi_{\text{part}}(t)}\ket{0}^{P}+
\sum\limits _{i}f_{i}(t)\ket{g_{2}}\hat{a}_{i}^{\dagger}\ket{0}^{P}
\end{eqnarray}
with $\ket{\psi_{\text{part}}(t)}$ constituting a linear superposition of the atomic states
$\ket{g_{1}}$, $\ket{g_{2}}$, $\ket{h}$, and $\ket{r}$. 
Taking into account the mode structure inside 
a parabolic cavity with large focal length \cite{Trautmann2014,Alber2013} 
the one photon probability amplitudes $f_{i}(t)$ can be eliminated
from the Schr\"odinger equation.
In the interaction picture we obtain the following effective inhomogeneous Schr\"odinger equation for the 
atomic state
\begin{eqnarray}
&&\iE\hbar\frac{d}{dt}\ket{\widetilde{\psi}_{\text{part}}(t)} = \hat{H}_{\text{part}}\ket{\widetilde{\psi}_{\text{part}}(t)}-|r\rangle\mathbf{E}_{\perp}^{\text{in},-}(\mathbf{x}_{a},t)
\cdot\mathbf{d}^{*},
\nonumber
\\
&&\hat{H}_{\text{part}}/\hbar = -\iE\frac{\Gamma}{2}|r\rangle\langle r|
-\label{eq:part_Schroedinger}\\
&&
\left(e^{\iE\Delta_{1}\left(t-t_{0}\right)}\frac{\Omega_{1}(t)}{2}|g_{1}\rangle\langle h|+
e^{\iE\Delta_{2}\left(t-t_{0}\right)}\frac{\Omega_{2}(t)}{2}|r\rangle\langle h|+\text{H.c.}\right).\nonumber
\end{eqnarray}
The term
\begin{eqnarray}
&\mathbf{E}_{\perp}^{\text{in}-}(\mathbf{x},t)=
\nonumber\\&
e^{\iE(\omega_{r}-\omega_{g_{2}})\left(t-t_{0}\right)}
\bra{\text{0}}\hat{\mathbf{E}}_{\perp}^{-}(\mathbf{x})
e^{-\frac{\iE}{\hbar}\hat{H}_{\text{field}}(t-t_{0})}\ket{\psi_{\text{field}}}\nonumber
\end{eqnarray}
describes the coherent driving of the atomic state $\ket{r}$ by
a freely evolving incoming single-photon wave packet. 
Similar to \cite{mollow1975pure} the  anti-Hermitian part of the Hamiltonian $\hat{H}_{\text{part}}$ takes into account
the spontaneous decay of the atomic state $\ket{r}$ .

For adiabatically eliminating the state $|h\rangle$ from the Schr\"odinger equation 
(\ref{eq:part_Schroedinger}) let us assume
that $\Delta_{1}=\Delta_{2}$ , that the Rabi frequencies are of the form
$\Omega_1(t)=\Omega\sin(\theta(t))$ and $\Omega_2(t)=\Omega\cos(\theta(t))$ ($\Omega>0$)
and that
the laser fields are sufficiently
intense \cite{incorporating_decay,bergmann1998coherent,STIRAP_decay}, i.e. $|\dot{\theta}(t)|,\Gamma\ll2\Omega^2/\mid\Delta_{1,2}\pm\sqrt{\Delta_{1,2}^2+4\Omega^2}\mid$.
Under these conditions
the adiabatic approximation applies \cite{raey} and  the atomic state $\ket{\psi_{\text{part}}(t)}$ can follow the dark
state $\ket{D(t)}=\cos(\theta(t))\ket{g_{1}}-\sin(\theta(t))\ket{r}$, i.e. $\ket{\widetilde{\psi}_{\text{part}}(t)}=c(t)\ket{D(t)}$. Using Eq.\,(\ref{eq:part_Schroedinger}) the amplitude $c(t)$ fulfills the relation
\begin{eqnarray}
\frac{d}{dt}c(t)=-\frac{\Gamma}{2}c(t)\sin(\theta(t))^{2}-\frac{\iE}{\hbar}\sin(\theta(t))\mathbf{E}_{\perp}^{\text{in},-}(\mathbf{x}_{a},t)\cdot\mathbf{d}^{*}\,.\nonumber\\
\label{Eq_diff_adiabatic}
\end{eqnarray}

In order to generate a time-symmetric single-photon wave packet in a readout process we start
from the initial states $\ket{\psi_{\text{atom}}}=\ket{g_{1}}$ and
 $\ket{\psi_{\text{field}}}=\ket{0}^{P}$ with
$\Omega_{2}(t_0)\gg\Omega_{1}(t_0)$ and $\sin(\theta(t_{0}))=0$.
In this case the last term in Eq. (\ref{Eq_diff_adiabatic}) does not contribute because the incoming photon field is in the vacuum state. Thus, the solution of Eq.(\ref{Eq_diff_adiabatic}) 
with the initial condition $c(t_0) = 1$ is given by
\begin{eqnarray}
c(t) &=& {\rm exp}\left(
-\frac{\Gamma}{2}
\int_{t_{0}}^{t}~dt^{\prime}
~\sin^2(\theta(t^{\prime}))
\right).
\label{c}
\end{eqnarray}
The envelope of the emerging single-photon wave packet
is determined by the atomic amplitude
$\braket{r|\widetilde{\psi}_{\text{part}}(t)}=-c(t)\sin(\theta(t))$. 
From Eq.(\ref{c})
one obtains the relation
\begin{equation}
\sin(\theta(t))=-\frac{\braket{r|\widetilde{\psi}_{\text{part}}(t)}}{\sqrt{1-\Gamma\int\limits _{t_{0}}^{t}|\braket{r|\widetilde{\psi}_{\text{part}}(t^{\prime})}|^{2}dt^{\prime}}}\,.
\label{eq:Loesung}
\end{equation}
Hereby, $\braket{r|\widetilde{\psi}_{\text{part}}(t)}$ has to be chosen in such a way that 
$\mid \sin(\theta(t))\mid \leq1$.
For the generation of a time-symmetric single-photon wave packet a possible choice of the atomic amplitude is given by
\begin{eqnarray}
\braket{r|\widetilde{\psi}_{\text{part}}(t)}&=&\left(\frac{2}{\pi\Gamma^{2}\sigma^{2}}\right)^{1/4}e^{-(t_{\text{max}}-t)^{2}/\sigma^{2}}
\end{eqnarray}
with $\sigma\Gamma$ and $(t_{\text{max}}-t_0)/\sigma$ being sufficiently large. 
The corresponding shapes of the STIRAP pulses $\Omega_1(t)$ and $\Omega_2(t)$
are determined by Eq. (\ref{eq:Loesung}).

Propagation of
the generated single-photon wave packet
to another far distant parabolic cavity through free space can be described 
by semi-classical methods \cite{Maslov1981}. Using the results
derived in Ref. \cite{Trautmann2014} 
the driving term entering Eq. (\ref{Eq_diff_adiabatic})
is determined by the quantum state of the first atom, i.e.
$\mathbf{E}_{\perp}^{\text{in},-}(\mathbf{x}_{2},t)\cdot\mathbf{d}^{*}=\alpha \Gamma \hbar
\braket{r|\widetilde{\psi}_{\text{part}}(t-\tau)}_{\text{atom 1}}$
with $\tau$ denoting
the time delay caused by photon propagation.
The complex valued constant $\alpha\in\mathbb{C}$  with $\mid \alpha \mid \leq 1$
describes a possible
phase shift and photon loss (with loss probability 
$1-|\alpha|^2$) during this propagation.

\begin{figure}[H]
\begin{center}
\includegraphics[width=0.39\textwidth]{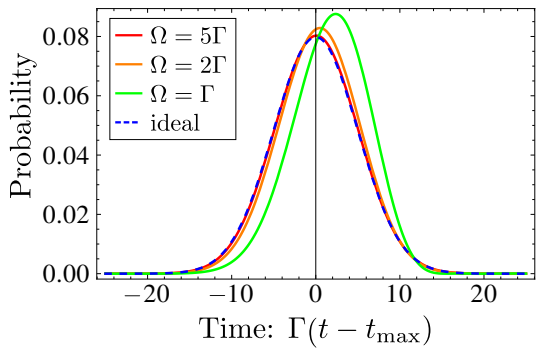}
\end{center}\caption{(color online). Time dependence of the probability  of detecting an atom in state $\ket{r}$ during the generation of 
the single-photon wave packet 
for several Rabi frequencies $\Omega$:
The parameters are $\sigma\Gamma=10$,  $\Gamma(t_0-t_{\text{max}})=25$ and $\Delta_1=\Delta_2=0\;.$
The result of the adiabatic approximation is denoted by 'ideal'. \label{fig:Probability_readout_r}}
\end{figure}

\begin{figure}[H]
\begin{center}
\includegraphics[width=0.39\textwidth]{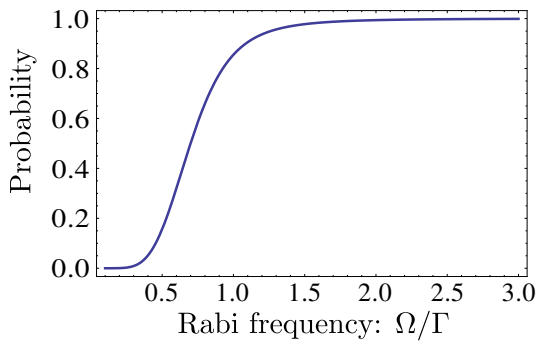}
\end{center}
\caption{(color online). Success probability of absorption of the single-photon wave packet generated by the readout procedure and its
dependence on the Rabi frequency $\Omega$ provided the photon is not lost during transmission (i.e. $\mid\alpha\mid=1$):
The parameters are $\sigma\Gamma=10$,  $\Gamma(t_0-t_{\text{max}})=25$ and $\Delta_1=\Delta_2=0$. \label{fig:Success_probability_storage}}
\end{figure}

In order to validate the results obtained from the adiabatic approximation we have solved
Eq.\,(\ref{eq:part_Schroedinger}) numerically.
The corresponding probability of finding the first atom in state $\ket{r}$ during the generation of the wave packet is depicted in 
Fig.\,\ref{fig:Probability_readout_r}.
These numerical results demonstrate that  a time-symmetric single-photon wave-packet can be tailored provided the 
Rabi frequency $\Omega$ is sufficiently high.
Another quantity of interest is the success probability for absorbing the wave packet generated by the first atom and
for transferring the population of the second atom from the initial state $\ket{g_{2}}$
to state $\ket{g_1}$ provided the photon is not lost during transmission, i.e. $\mid \alpha \mid = 1$. This probability is depicted in
Fig.\,\ref{fig:Success_probability_storage}. Hereby we assume that the readout and retrieval
processes are performed with the same Rabi frequency $\Omega$. With increasing $\Omega$ this probability quickly approaches unity.
Thus, our scheme leads to almost perfect absorption of
the wave packet if $\Omega$ is large in comparison with the spontaneous decay rate $\Gamma$.
Furthermore, numerical calculations confirm that the maximal probability of finding the atoms in state $\ket{h}$
scales with
$\Omega^{-2}$. For $\Omega=10\Gamma$ and the parameters chosen in Fig. \ref{fig:Probability_readout_r} and Fig. \ref{fig:Success_probability_storage}, for example,
its values are smaller than $10^{-3}$. This is valid for STIRAP-assisted
photon emission as well as absorption.
Thus, the assumption of a negligible population of the ancillary atomic level
$\ket{h}$
can be satisfied easily for sufficiently high Rabi frequencies.
\begin{figure*}
\begin{center} \includegraphics[scale=0.45]{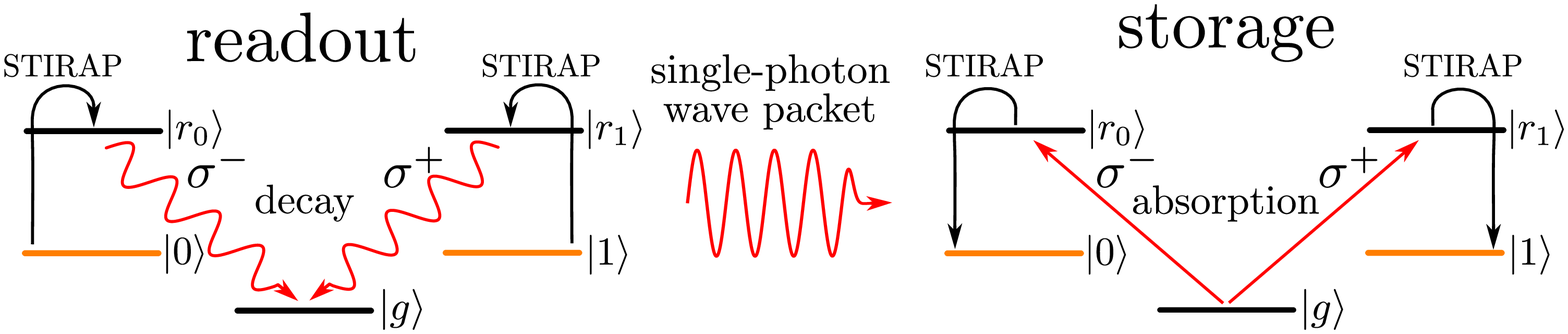}\end{center} 
\caption{\label{fig:advanced_state_transfer}(color online). Advanced protocol for quantum state transfer in free space:
The qubit states are denoted by  $\ket{0}$ and $\ket{1}$ (colored in orange). The nondegenerated ground state of the atom is denoted by $\ket{g}$. The red arrows indicate the relevant spontaneous
decay and photon absorption processes. The black arrows indicate the relevant STIRAP processes.
}
\end{figure*}

Based on these results quantum state transfer protocols suitable for free-space communication can be constructed.
In the simplest protocol the previously discussed level structure is used
for encoding a qubit in the atomic levels
$|g_{1}\rangle$ and $|g_{2}\rangle$ of an atom trapped in the focus of a parabolic cavity. The readout of this qubit
is performed by
the proposed STIRAP-assisted spontaneous photon emission process. Thus, if
this atom is prepared initially in state $|g_{1}\rangle$ it
is transferred to
state $|g_{2}\rangle$ after this spontaneous photon emission process. 
Alternatively, if it is prepared initially
in state $|g_{2}\rangle$ it remains in this state and
no photon is emitted. Correspondingly, if initially the atom is in a linear superposition of both states
this qubit
state is coherently mapped onto a 
coherent superposition of the presence and absence of a single photon. After photon propagation through free space the retrieval of this photonic qubit is accomplished
by the time reversed STIRAP-assisted process acting on a second atom initially prepared in
state $|g_{2}\rangle$. This way it is possible to 
store the original qubit again in a second distant matter qubit positioned in the focal point of another parabolic cavity.
However, this simple procedure has the drawback that photon
loss cannot be detected by the receiver  and leads to a significant reduction of the fidelity of the original qubit state.

This problem can be circumvented by an advanced free-space quantum-communication protocol whose atomic excitation scheme is depicted in Fig.\,\ref{fig:advanced_state_transfer}. In this protocol
the photonic qubit is encoded
in the polarization state of the generated single-photon wave packet and the matter qubit may be formed by Zeeman sublevels of a degenerate metastable atomic state.
After photon propagation the retrieval of this photonic qubit and storage in another matter qubit is achieved with the corresponding time-reversed STIRAP assisted
single-photon absorption process.
In this protocol photon loss only affects the success probability of the retrieval process and not the fidelity of the retrieved qubit state.
Provided the transmission of the generated single-photon wave packet is successful
the quantum state of the first atom is transferred to the distant
second atom with fidelity arbitrarily close to unity if the Rabi frequency $\Omega$ is sufficiently large.
Typical experimental
imperfections, such as finite sizes of the parabolas or real surface properties reduce the success probability but leave the fidelity unaltered \cite{Trautmann2014}.

Atomic level structures suitable for implementing such a scheme are available in
alkaline-earth and alkaline-earth-like atoms, for example. These atoms offer ground states with vanishing electronic spin and are therefore of interest for quantum information processing \cite{PhysRevLett.101.170504,PhysRevLett.102.110503}
and for realizations of optical frequency standards \cite{Katori2003}.
For this scheme isotopes with vanishing nuclear spin are of particular interest due to their nondegenerate ground states.
The matter qubit may be encoded in long-lived metastable states and  the single-photon wave packet may be generated
by dipole allowed optical transitions from suitable excited states to the nondegenerate ground state of the atom.
The presence of such a nondegenerate ground state is necessary in order to
ensure that the excited states decay to only one state of the level structure.
Successful cooling and trapping of such atoms, such as 
${}^{88}\text{Sr}$ and ${}^{174}\text{Yb}$, has already been reported 
\cite{trapping_strontium,PhysRevLett.91.040404}. In these neutral atoms also suitable level schemes can be selected \cite{Possible-Levels,leveldata}.

In conclusion, we have proposed  a STIRAP-assisted method for generating time-symmetric single-photon wave packets on demand by spontaneous decay in free space. 
With the help of these single-photon wave packets high-fidelity quantum state transfer 
between distant matter qubits by photon propagation in free space can be implemented.
In contrast to already known quantum state transfer schemes our protocol does not require high-finesse cavities and optical fibers so that it offers interesting perspectives for applications
in free-space quantum communication.

\begin{acknowledgments}
  Stimulating discussions with G. Birkl are acknowledged.
  This work is supported by CASED III, DAAD, the BMBF Project Q.com, and by the DFG as part of the CRC 1119 CROSSING.
  G.S.A thanks the wonderful hospitality of MPL, Erlangen.
\end{acknowledgments}


\begin{thebibliography}{24}
\bibitem{incorporating_decay}
  The spontaneous decay rate $\Gamma_h$ of level $\ket{h}$ can be incorporated in this condition by choosing $\text{Im}[\Delta_{1,2}]=\Gamma_{h}/2$.
\bibitem{Possible-Levels}

A possible level scheme in ${}^{88}\text{Sr}$ is:
$\ket{g}\equiv 5s^{2}{\;}^{1}S_{0}$;
$\ket{0},\ket{1}\equiv 5s4d{\;}^{1}D_{2}$ ($m=\pm1$);
$\ket{r_0},\ket{r_1}\equiv 5s5p{\;}^{1}P_{1}$ ($m=\pm1$);
the STIRAP levels $\ket{h_0},\ket{h_1}\equiv 5s4f{\;}^{1}F_{3}$ ($m=\pm1$).
The laser-induced transfer between levels $\ket{h_0}$, $\ket{h_1}$ and $\ket{r_0}$, $\ket{r_1}$ involves a two-photon process (e.g. via the intermediate levels $5s5d{\;}^{1}D_{2}$ ).
\end{thebibliography}
\end{document}